\documentclass[12pt,preprint]{aastex}
\makeindex
\makeatletter
\usepackage{booktabs}

\title{Testing nonstandard cosmological models with SNLS3 supernova data and other cosmological probes}

\author{Zhengxiang Li$^{a}$, Puxun Wu$^{b, c}$  and Hongwei Yu$^{a, b, c, } \footnote{Corresponding author: hwyu@hunnu.edu.cn}$ }

\affil{$^a$Department of Physics and Key Laboratory of Low
Dimensional Quantum Structures and Quantum Control of Ministry of
Education, Hunan Normal University, Changsha, Hunan 410081, China
\\$^b$Center of Nonlinear Science and Department of Physics, Ningbo
University,  Ningbo, Zhejiang 315211, China \\
$^c$Kavli Institute of Theoretical Physics China, CAS, Beijing
100190, China}

\begin{abstract}
We investigate the implications for some nonstandard cosmological
models using data from the first three years of the Supernova Legacy
Survey (SNLS3), assuming a spatially flat universe. A comparison
between the constraints from the SNLS3 and those from other SN Ia
samples, such as the ESSENCE, Union2, SDSS-II and Constitution
samples, is given and the effects of different light-curve fitters
are considered. We find that SN Ia with SALT2 or SALT or SIFTO can
give consistent results and the tensions between different data sets
and different light-curve fitters are obvious for
fewer-free-parameters models. At the same time, we also study the
constraints from the SNLS3 along with data from the cosmic microwave
background and the baryonic acoustic oscillations (CMB/BAO), and the
latest Hubble parameter versus redshift (H(z)). Using model
selection criteria such as $\chi^2$/dof, GoF, AIC and BIC, we find
that, among all the cosmological models considered here
($\Lambda$CDM, constant $w$, varying $w$, DGP, modified polytropic
Cardassian, and the generalized Chaplygin gas), the flat DGP is
favored by the SNLS3 alone. However, when additional CMB/BAO or H(z)
constraints are included, this is no longer the case, and the flat
$\Lambda$CDM becomes preferred.
\end{abstract}


\begin{document}

\maketitle

\section{INTRODUCTION}
Type Ia supernovae (SNe Ia)  have altered the focus of cosmology
dramatically since they first indicated  that the expansion of the
Universe is currently accelerating~\citep{Riess, Perlmutter}.
Usually, the mysterious cause of the cosmic acceleration is
attributed either to the existence of an exotic energy component,
called dark energy, or a modification of  the standard theory of
gravity. At present, SNe Ia are still the most direct probe of the
history of cosmic expansion and the properties of dark energy, and
many SN Ia samples, such as ESSENCE~\citep{Wood-Vasey},
SDSS-II~\citep{Kessler}, Constitution~\citep{Hicken} and
Union2~\citep{Amanullah}, have been released.

The Supernova Legacy Survey (SNLS) is an ongoing five year project
that aims to probe the expansion history of the universe using SNe
Ia~\citep{Sullivan2005}. The goal of this survey is to measure the
time-averaged equation of state (EOS) of dark energy ($w$) to $5\%$
(statistical uncertainties only) in combination with other
cosmological probes and to $10\%$ including the systematic
uncertainties. Analysis of the first year of data from SNLS,
containing 71 high-redshift SNe Ia, was presented in~\citet{Astier}.

Recently, data from the first three years  SNLS data set were
released. \citet{Guy} presented the photometric properties and
relative distance moduli of 252 SNe Ia from this sample. Based on
selection criteria, they removed 21 SNe and obtained
$\Omega_m=0.211\pm0.034(\mathrm{stat})\pm0.069(\mathrm{sys})$ for a
flat $\Lambda$CDM using the SNLS3 data alone. Using a slightly
different set of selection criteria, \citet{Conley} selected 242 SNe
Ia from SNLS, and combined these with 123 low-$z$, 93
intermediate-$z$ (SDSS-II) and 14 high-$z$ SNe from the
{Hubble~Space~Telescope} to form a high-quality unified sample
containing 472 SNe Ia. We shall refer to this sample as SNLS3. For
the SNLS3 data alone, C11 finds that
$w=-0.91^{+0.16}_{-0.20}(\mathrm{stat})^{+0.07}_{-0.14}(\mathrm{sys})$,
considering a flat Universe with constant $w$. The constraints on
dark energy  when SNLS3 is combined with other probes (such as the
baryonic acoustic oscillation (BAO)~\citep{Percivalb} or
WMAP7~\citep{Komatsub}) were presented in~\citet{Sullivan}. The
results for a flat constant $w$ universe are $\Omega_{m} = 0.269 \pm
0.015$ and $w =-1.061^{+0.069}_{-0.068}$, and those for a flat
varying $w$ universe characterized by the CPL
parameterization~\citep{CPL1, CPL2} ($w(a) = w_0 + w_a(1-a)$)  are
$w_0 = -0.905\pm 0.196$ and $w_a =-0.984^{+1.094}_{-1.097}$.

Dark energy is not the only possible explanation of the present
cosmic acceleration, and many other models have been considered such
as DGP, Cardassian, and the generalized Chaplygin gas. In this paper
we investigate constraints on these nonstandard cosmological models.
Similar analyses were carried out by \citet{Davis} and
~\citet{Sollerman} using the ESSENCE~\citep{Wood-Vasey} and
SDSS-II~\citep{Kessler} SN Ia data sets, respectively. In order to
give a comparison between the results from SNLS3 with those from
other SN Ia data sets, we also compare the constraints obtained
using SNLS3 to those using other popular SNe Ia samples such as the
ESSENCE, Constitution, Union2 and SDSS-II samples, and additionally
analyze the effects of different light-curve fitters on the results.
In addition to the SNe Ia data, constraints from other probes, such
as the cosmic microwave background (CMB) and baryonic acoustic
oscillations (BAO) measurements of the acoustic scale (CMB/BAO) and
the Hubble parameter vs. redshift (H(z)), are also considered.

We organize our paper as follows. Section 2 presents the data sets
and the statistical analysis method. Section 3 describes briefly the
models considered and the observational constraints on them. Model
testing results using model selection statistics are shown in
section 4. Finally, Section 5 gives a summary of our results.

\section{DATA SETS AND ANALYSIS METHODS}
In this Section, we describe the data sets and analysis techniques.

\subsection{Type Ia supernovae}
The SNLS3 SN Ia sample has been discussed in detail in C11. This
sample consists of 472 SNe Ia (123 low-$z$, 93 SDSS, 242 SNLS, and
14 $Hubble~Space~Telescope$) and is compiled with the combined SiFTO
\citep{Conley08} and SALT2 \citep{Guy07} light curve fitters. As
discussed in S11, the  SNLS3 data set has several advantages over
the first year sample presented in \citet{Astier}. First of all, the
sample size has increased by a factor of three. Furthermore, the
sources of potential astrophysical systematics are examined by
dividing the SN Ia sample according to the properties of either the
SN or its environment \citep{Sullivan10}. Finally, an ameliorated
photometric calibration of the light curves and a more consistent
understanding of the experimental characteristics are
allowed~\citep{Regnault}.

For the cosmological analysis, one can minimize the following
$\chi^2$,
\begin{equation}
\chi^2=\sum_{\mathrm{SN}}\frac{(m_B-m_B^{mod})^2}{\sigma^2_{stat}+\sigma^2_{int}},
\end{equation}
where $m_B$ is the rest-frame $B$-band magnitude of an SN,
$m_B^{mod}$ is the magnitude of the SN predicted by the cosmological
model,  $\sigma_{stat}$ includes the uncertainties both in $m_B$ and
$m_B^{mod}$ and  $\sigma_{int}$ parameterizes the intrinsic
dispersion of each SN sample. The values for $\sigma_{int}$ used in
our analysis can be found  in Table 4 in C11.  The model-dependent
magnitude is given by
\begin{equation}
m_B^{\mathrm{mod}}=5\log_{10}\mathcal{D}_L(z_{\mathrm{hel}},z_{\mathrm{cmb}};\mathbf{p})-\alpha(s-1)+\beta\mathcal{C}+\mathcal{M}_B,
\end{equation}
where $\mathbf{p}$ stands for the model parameters set,
$\mathcal{D}_L$ is the Hubble-constant free luminosity distance,
$z_{\mathrm{hel}}$ and $z_{\mathrm{cmb}}$ are the heliocentric and
CMB frame redshifts of the SN, and $\alpha$ and $\beta$ characterize
the stretch and color ($s$ and $\mathcal{C}$)-luminosity
relationships. $\mathcal{M}_B$ represents some combination of the
absolute magnitude of a fiducial SN Ia and the Hubble constant.
Here, $\alpha$, $\beta$ and $\mathcal{M}_B$ are ``nuisance
parameters". We marginalize over $\mathcal{M}_B$ by following the
method described in C11 (Appendix C). In this marginalization method
the environmental dependence of SN properties, which allows
$\mathcal{M}_B$ to be split by host-galaxy stellar mass at
$10^{10}M_\odot$ (this correction is not employed for all other SN
Ia samples we are considering), is taken into account. Defining the
vector of residuals between the model magnitudes and the observed
magnitudes
$\Delta\mathbf{\vec{m}}=\mathbf{\vec{m}}_B-\mathbf{\vec{m}}_B^{\mathrm{mod}}$,
we obtain the observational constraints on model parameters by
minimizing
\begin{equation}
\chi^2_{SN}=\Delta\mathbf{\vec{m}}^T\cdot\mathbf{C}^{-1}\cdot\Delta\mathbf{\vec{m}},
\end{equation}
where $\mathbf{C}^{-1}$ is the inverse covariance
matrix~\footnote{https://tspace.library.utoronto.ca/handle/1807/26549}.
The total covariance matrix is detailed in section 3.1  in C11. The
covariances which take both statistical and systematic uncertainties
into consideration are used in our investigation. We allow $\alpha$
and $\beta$ to vary  with the cosmological parameters to avoid
biasing our results using a grid $\chi^2$ minimization routine and
then use their best fit results to obtain the allowed regions of
model parameters.

It is well-known  that to obtain the precise distance of the
standard candle, SNe Ia, is crucial for modern cosmology. Since the
intrinsic luminosity of SNe Ia is correlated with the shape of its
optical light curves, one can determine it by proposing a method to
relate them. \citet{Phillips1993} firstly introduced a method by
finding a correction between the SNe Ia intrinsic luminosity and the
parameter $\Delta m_{15}(B)$, where $\Delta m_{15}(B)$ is the amount
of a SNe Ia B-band  declination during the first fifteen days after
maximum light. The MLCS is a different method using the multicolor
light curve shapes to estimate the luminosity distance proposed in
\citet{Riess95}.  This approach was extended to include U-band
measurements, named MLCS2k2,  by \citet{Jha2002}.  The MLCS2k2  has
been applied for distance estimate in several popular SN Ia samples,
such as ESSENCE, SDSS-II, and Constitution.

\citet{Guy05} proposed another method, SALT, to construct the SNe Ia
luminosity distance estimate by parameterizing the light curve with
a parameter set, i.e.,  a luminosity parameter, a decline rate
parameter and a single color parameter. It offers a practical
advantage which makes it easily applicable to high-redshift SNe Ia.
By including spectroscopic data, \citet{Guy07} improved this SALT to
SALT2. SALT2 has been applied to calculate the distance modulus for
several popular SN Ia samples, such as Constitution, Union2,
SDSS-II, and SNLS3. It is notable that the SDSS-II corrected the
SALT2 results for selection biases by using a Monte Carlo simulation
(see section 5.2 and section 6 of Kessler et al. 2009) and the SNLS
team made a few technical modifications in the training procedure,
such as higher resolution for the components and the color variation
law, and a new regularisation scheme, to SALT2 for the SNLS3
(detailed in section 4.3 and Appendix A of Guy et al. 2010).
Recently, \citet{Conley08} proposed another new empirical way, i.e.,
SiFTO, which is similar to SALT2 using the luminosity magnitude,
light curve shape and color to obtain the distance. This method has
been applied in the analysis of SNLS3.

Previous works have shown that different SN Ia data sets give
different results \citep{Rydbeck, Bueno} and different light-curve
fitters may also lead to different conclusions although the same SN
Ia sample is considered~\citep{Sollerman, Bengochea}. Therefore, in
the present paper, besides the SNLS3 (analyzed with SALT2 and SiFTO
light-curve fitters), we also consider some other popular SN
compilations: ESSENCE (SALT and MLCS2k2), Constitution (SALT2 and
MLCS2k2 with $R_V=1.7$), Union2 (SALT2) and SDSS-II (SALT2 and
MLCS2k2). However, it should be mentioned that these SN Ia samples
are not independent, since many of them are drawn from the same
sources. That is , different SN Ia compilations may contain many of
the same SNe. For instance, the nearby~\citep{Jha} and
SNLS~\citep{Astier} SN Ia are, in part, included as subsets in all
above mentioned samples. In addition, except for the Union2 and
SNLS3, there is no easy way to include systematic uncertainties for
other referred SN samples. In our analysis only for the SNLS3 sample
are  the systematic uncertainties considered.

\subsection{The Cosmic Microwave Background  and Baryon Acoustic Oscillations}
The CMB and BAO constraints used in this analysis are based on the
angular scales measured at the CMB decoupling epoch at $z_*=1090$
and imprinted in the clustering of luminous red galaxies (BAO) at
$z=[0.2,0.35]$. Its acoustic scale is given by
\begin{equation}
l_A=\pi \frac{d_A(z_*)(1+z_*)}{r_s(z_*)},
\end{equation}
which represents the angular scale of sound horizon at decoupling,
where $d_A=d_L/(1+z)^2$ is the proper (not comoving) angular
diameter distance, $d_L=(1+z)\int_0^z\frac{dz}{H}$ is the luminosity
distance,  and $r_s(z_*)$ is the comoving sound horizon at
recombination,
\begin{equation}
r_s(z_*)=\int_{z_*}^\infty \frac{c_s(z)}{H(z)}dz
\end{equation}
which depends on the speed of sound, $c_s$, in the early universe.
The BAO scale is given by $r_s(z_d)/D_V$, where the $r_s(z_d)$ is
the comoving sound horizon at drag epoch ($z_d\approx1020$), and the
so-called dilation scale, $D_V$, is a combination from angular
diameter distance and radial distance according to
\begin{equation}
D_V(z)=\bigg[(1+z)^2d_A^2\frac{cz}{H(z)}\bigg]^{1/3}.
\end{equation}

By combining the BAO measurements of $r_s(z_d)/D_V(z)$ at two
redshifts~\citep{Percivala}, $r_s(z_d)/D_V(0.2)=0.1980\pm0.0058$ and
$r_s(z_d)/D_V(0.35)=0.1094\pm0.0033$, with the CMB measurement of
$l_A$ given in~\citet{Komatsua}, $l_A=302.10\pm0.86$, and
considering the ratio of the sound horizon at the two epochs,
$r_s(z_d)/r_s(z_*)=1.044\pm0.019$, the final constraints we use for
the cosmology analysis are obtained:
\begin{equation}
f_{0.20}=\frac{d_A(z_*)(1+z_*)}{D_V(z=0.2)}=19.04\pm0.58
\end{equation}
\begin{equation}
f_{0.35}=\frac{d_A(z_*)(1+z_*)}{D_V(z=0.35)}=10.52\pm0.32,
\end{equation}
which do not depend on the comoving sound horizon scale at
recombination. However, these two ratios are not independent with a
correlation coefficient of $0.39$. This correlation is considered in
our cosmological fits. Moreover, the constraints from these two
ratios (labeled CMB/BAO here) are expected to be a good
approximation for all models tested in this paper since the redshift
difference between the decoupling and the drag epoch is relatively
small, and the sound horizon at these two epoch is mostly dominated
by the fractional difference between the number of photons and
baryons~\citep{Sollerman}.

In addition, we do not use the CMB ``shift parameters" ($l_A, R,
z_*$) derived from WMAP7~\citep{Komatsub} in our analysis, because
the CMB distance prior is applicable only when the model in question
is based on the standard Friedmann-Lemaitre-Robertson-Walker
universe with matter, radiation, dark energy and spatial
curvature~\citep{Komatsub}, and we are testing some nonstandard
models, e.g. DGP and modified polytropic Cardassian.

\subsection{Hubble parameter versus redshift data}
The Hubble parameter versus redshift data to be used in this paper
consists of three subsamples. The first one includes 9 data points
which are derived from differential ages of old passively evolving
galaxies~\citep{simon}. The second two-points subsample is
determined from the high-quality spectra with the Keck-LRIS
spectrograph of red-envelope galaxies in the redshift range
$0.2<z<1$~\citep{Hubble1}. The third subsample contains three data
points obtained by using the 2-point correlation of SDSS luminous
red galaxies and taking the BAO peak position as a standard ruler in
the radial direction~\citep{Hubble3}. These three points are
considered as a direct measurement of H(z) for the first time and
are independent of the earlier BAO measurements~\citep{Percivala}
which  constrains an integral of H(z) by using the spherically
averaged (monopole) correlation. The parameters set can be
determined by minimizing
\begin{equation}{\label{Eq1}}
\chi^2_{\mathrm{H(z)}}=\sum_{i=1}^{14}\frac{[H_{th}(z_i;\mathbf{p})-H_{obs}(z_i)]^2}{\sigma^2_i}.
\end{equation}
When fully expanded, this expression includes $H_0$ as a nuisance
parameter. We marginalize over $H_0$ using the Gaussian prior
$H_0=73.8\pm2.4$ $km~s^{-1}~Mpc^{-1}$~\citep{Hubble2} as proposed in
\citep{puxun}.

\section{MODELS AND CONSTRAINT RESULTS}
For the sake of  simplicity, we assume a  spatially flat universe.
We study several popular cosmological models. The chosen models,
their parameters, and the abbreviations we use to refer to them are
summarized in Table~\ref{Tab1}. In the following, we will outline
the basic equations governing the background evolution of the
universe in each
 model and give the observational results.

\begin{table}[!h]
\begin{center}
\begin{tabular}{l}
\toprule[1.5pt]
 ~~~~~~~~~~~Model~~~~~~~~~~~~~~~~~~~~~~~~~~~~~~~~~~~~~~~~~~Abbreviation~~~~~~~~~~~~~~~~~~~~Parameters~\\
\midrule[0.5pt]
 Flat $\Lambda$CDM.................................................~~~~~~~~~~~F$\Lambda$~~~~~~~~~~~~~~~~~~~~~~~~~~~~~~~~~$\Omega_m$\\
 Flat constant $w$...........................................~~~~~~~~~~~F$w$~~~~~~~~~~~~~~~~~~~~~~~~~~~~~~$\Omega_m,~w$\\
 Flat varying $w$ (CPL)..................................~~~~~~~~~FCPL~~~~~~~~~~~~~~~~~~~~~~~~~~$\Omega_m,~w_0,~w_1$\\
 Flat Dvali-Gabadadze-Porrati brane ..........~~~~~~~~~FDGP~~~~~~~~~~~~~~~~~~~~~~~~~~~~~~~$\Omega_{r_c}$\\
 Flat Modified Polytropic Cardassian..........~~~~~~~~~~FMPC~~~~~~~~~~~~~~~~~~~~~~~~~~$\Omega_m,~q,~n$\\
 Flat Generalized Chaplygin Gas..................~~~~~~~~~FGCG~~~~~~~~~~~~~~~~~~~~~~~~~~~~~$A_s,~\gamma$\\
\bottomrule[1.5pt]
\end{tabular}
\end{center}
\caption{\label{Tab1} Summary of models. $\mathrm{\mathbf{Note}}$:
since the parameter ``$\alpha$" has been used to describe the
stretch-luminosity relation for the SNLS3, we utilize ``$\gamma$" to
replace the original parameter ``$\alpha$" in the GCG model.}
\end{table}

\subsection{Flat $\Lambda$CDM}
\begin{figure}%[h!]
\centering
\includegraphics[width=0.8\linewidth]{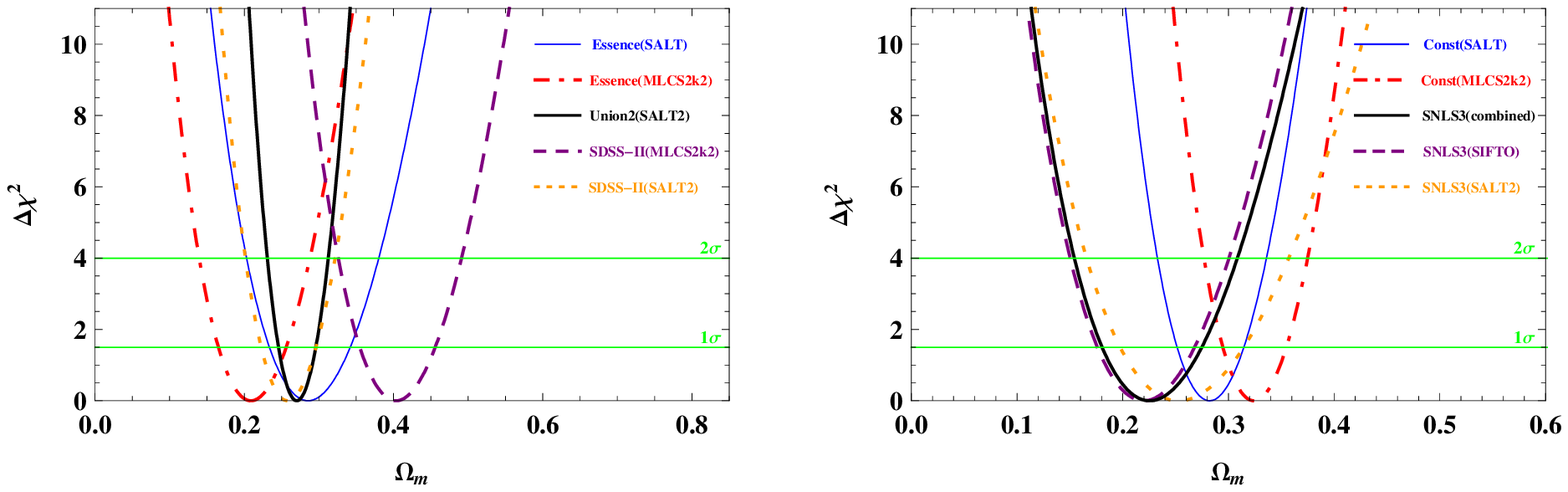}
\caption{ The constraints on the
flat $\Lambda$CDM model from several popular SN Ia data sets
compiled with different light-curve fitters. ``Const" represents for
the Constitution SN Ia sample.\label{FigL1}}
\end{figure}

\begin{figure}[h!]
   \centering
       \includegraphics[width=0.45\linewidth]{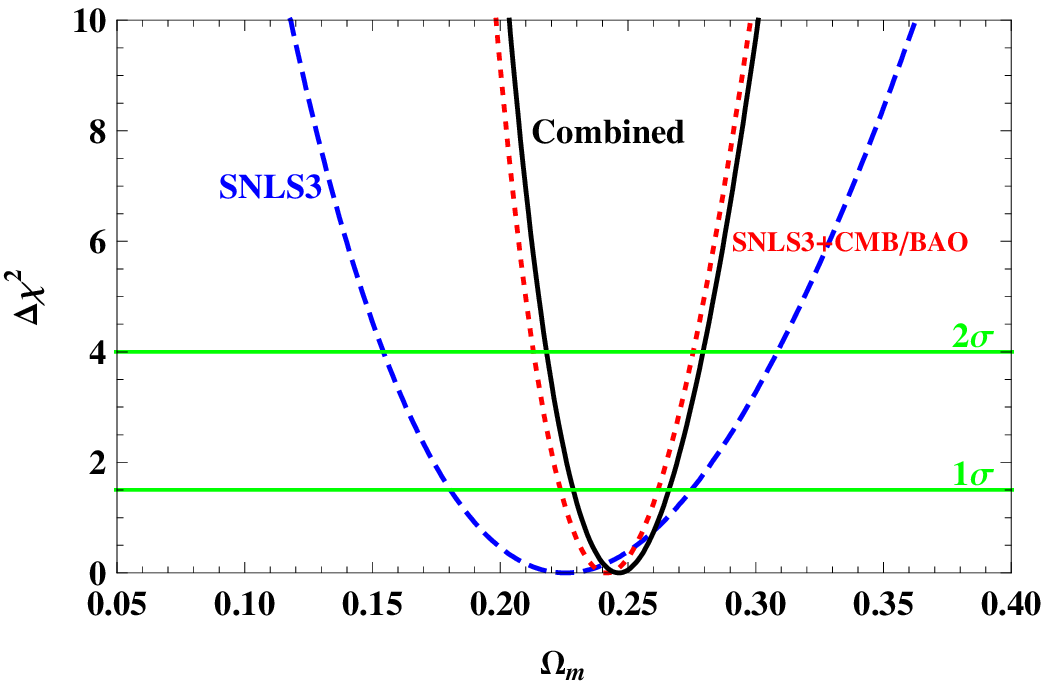}
   \caption{\label{FigL2} The constraints on the
flat $\Lambda$CDM model. The blue dashed, red dotted and solid lines
represent the results from SNLS3, SNLS3+CMB/BAO and
SNL3+CMB/BAO+H(z), respectively. }
\end{figure}

The flat $\Lambda$CDM model is the simplest one that can explain the
present accelerating cosmic expansion, and is generally considered
to be the standard cosmological model.  In this model, dark energy
is a result of a nonzero  cosmological constant. So, we have
$\Omega_{\mathrm{DE}}=\Omega_\Lambda=(1-\Omega_m)$, with equation of
state $w=-1$. The Friedmann equation in this case is
\begin{equation}
H^2=H_0^2[\Omega_m(1+z)^3+(1-\Omega_m)],
\end{equation}
This model is consistent with almost all observations.
Fig.~\ref{FigL1} shows a comparison of different SN Ia data
constraints on the model parameter.  From this Figure, we see that
irrespective of whether SALT2, SALT, or SIFTO light curve fitter is
used, different SN Ia samples give fairly consistent results.
However, this is not the case when the MLCS2k2 light-curve fitter is
considered, different SN Ia sets lead to notably different results.
For example, the best fit values are
$\Omega_m=0.209^{+0.059}_{-0.054}$ (ESSENCE),
$\Omega_m=0.324^{+0.038}_{-0.036}$ (Constitution) and
$\Omega_m=0.403^{+0.056}_{-0.059}$ (SDSS-II).  In addition,
as~\citet{Sollerman}, we also find that SDSS-II (SALT2) and SDSS-II
(MLCS2k2) are clearly inconsistent. At the $95.4\%$ confidence
level, the constraints from SDSS-II (SALT2) and SDSS-II (MLCS2k2)
have no overlap.  In Fig.~\ref{FigL2}, we give the constraints from
SNLS3 with other cosmological probes, from which we find that the
main constraints come from the SN Ia+CMB/BAO and the H(z) data set
has little effect. Using SNLS3+CMB/BAO+H(z), we obtain
$\Omega_m=0.245^{+0.026}_{-0.020}$ at the $68.3\%$ confidence level.
Compared with the value of $\Omega_m = 0.27 \pm 0.04$ obtained by
Davis et al. (2007) from ESSENCE+BAO+CMB, we find that
SNLS3+CMB/BAO+H(z) yields a tighter constraint on $\Omega_m$.

\subsection{Flat constant $w$}\label{section 3.2}
\begin{figure}[h!]
   \centering
       \includegraphics[width=0.8\linewidth]{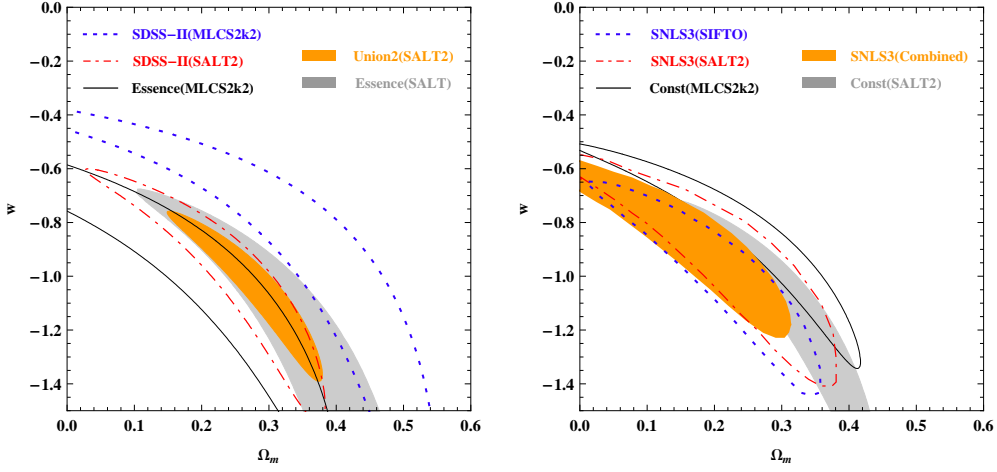}
   \caption{\label{Figw1} The $68.3\%$ contours for the
flat constant $w$ model from several popular SN Ia data sets
compiled with different light-curve fitters. ``Const" represents for
the Constitution SN Ia sample.}
\end{figure}

\begin{figure}[h!]
   \centering
       \includegraphics[width=0.45\linewidth]{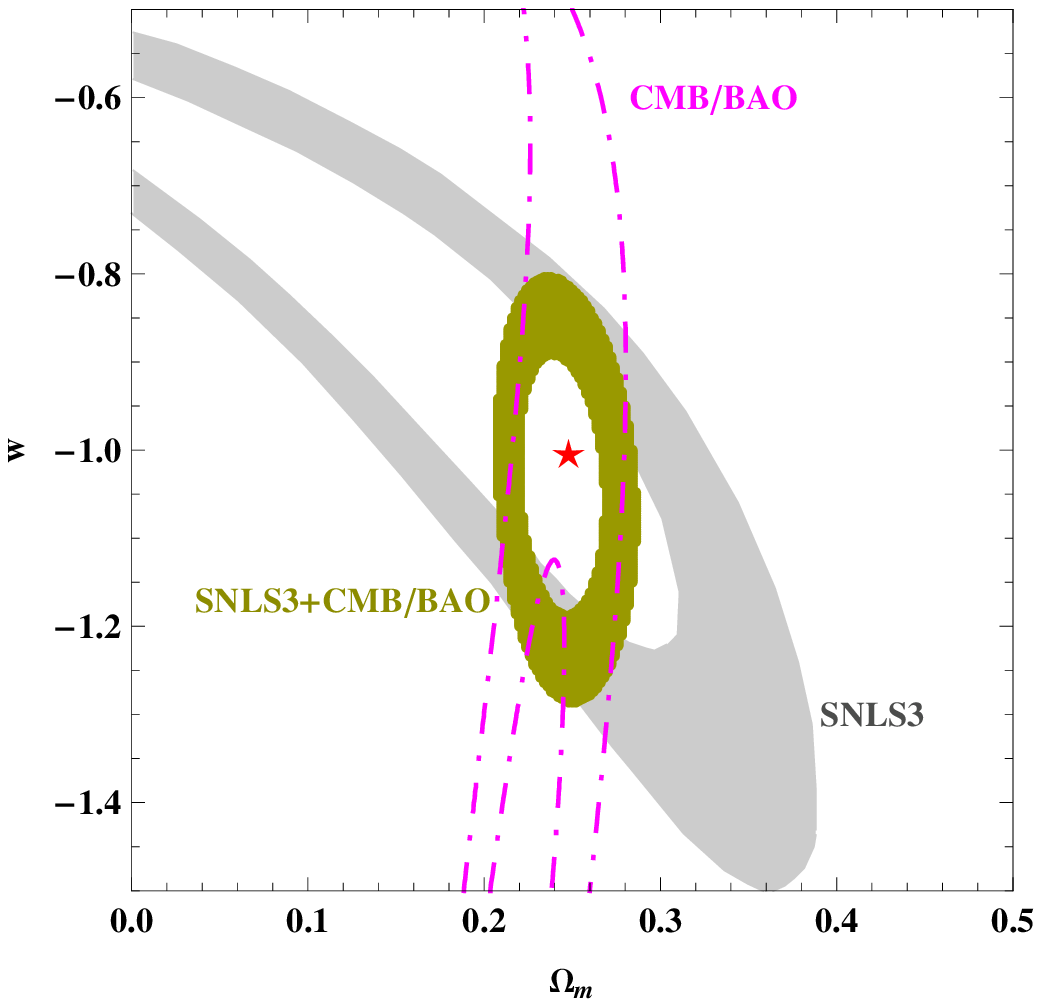}
       \includegraphics[width=0.45\linewidth]{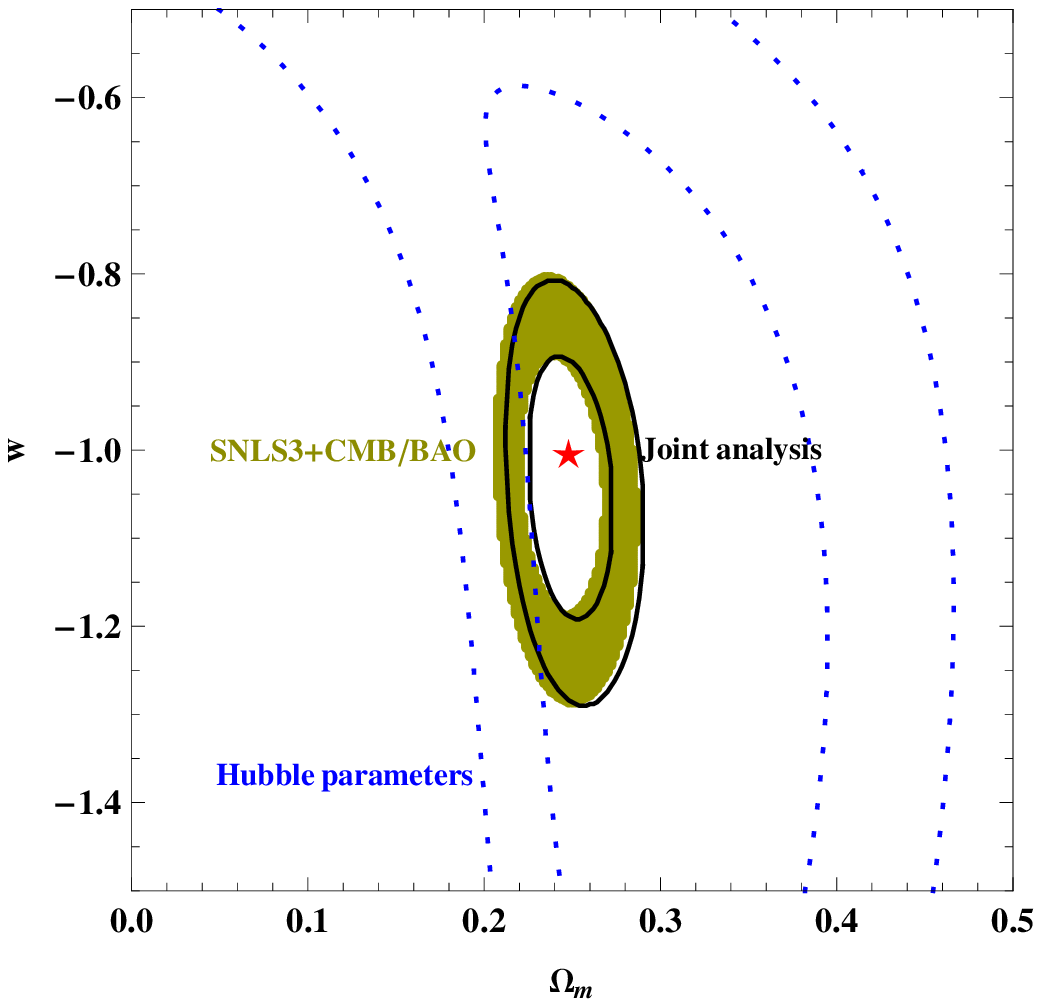}
   \caption{\label{Figw2} The $68.3\%$ and $95.4\%$ confidence level contours for
the flat constant $w$ model.  Left panel shows the constraints from
SNLS3, CMB/BAO and their combination,  and right panel  is the case
with the H(z) data included.
   The red star ($\Omega_m=0.25,~w=-1.0$)
   represents the $\Lambda$CDM with the best fit $\Omega_m$}
\end{figure}

This model is obtained by assuming that dark energy has a constant
equation of state parameter $w$. Thus the Friedmann equation can be
expressed as:
\begin{equation}
H^2=H_0^2[\Omega_m(1+z)^3+(1-\Omega_m)(1+z)^{3(1+w)}],
\end{equation}
which depends on two free parameters, $\Omega_m$ and $w$.
Fig.~\ref{Figw1} shows the constraints at the $68.3\%$ confidence
level from different SN Ia samples, from which we find that, for all
SN Ia data, especially for the SDSS-II, different light-curve
fitters still affect the results. In the SDSS-II case, there exists
an obvious inconsistency between SALT2 and MLCS2k2 since there is no
overlap at the $68.3\%$ confidence level in this model. By fitting
this model to the combined SNLS3 SN Ia, BAO/CMB and H(z) data, the
$68.3\%$ constraints are $\Omega_m=0.248^{+0.024}_{-0.022}$ and
$w=-1.039^{+0.135}_{-0.153}$. The contours of $\Omega_m$ and $w$ are
plotted in Fig.~\ref{Figw2}. $w=-1$ is consistent with observations
at the $68.3\%$ confidence level. As was the case for the
$\Lambda$CDM, the H(z) has little effect on the derived cosmological
parameters.

\subsection{Flat varying $w$ (CPL)}
\begin{figure}[h!]
   \centering
       \includegraphics[width=0.8\linewidth]{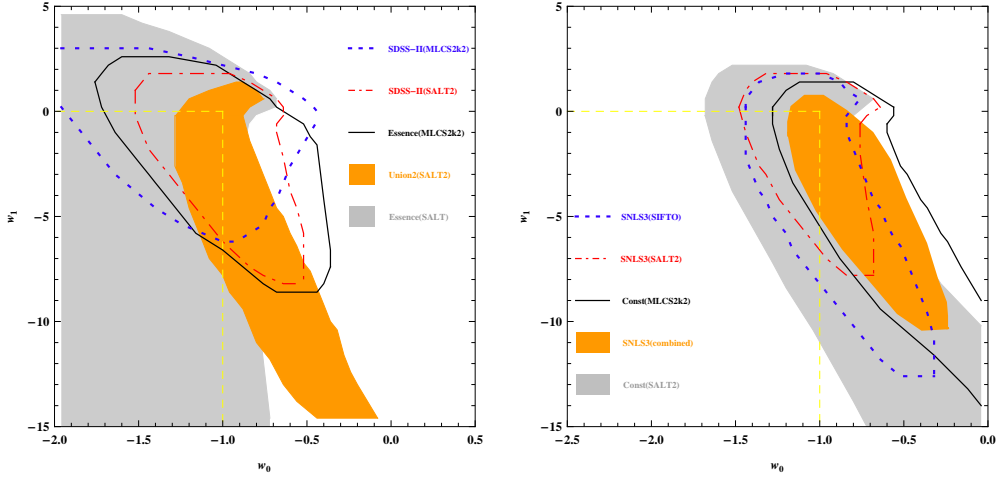}
   \caption{\label{FigC1}  The marginalized $68.3\%$ contours of $w_0$ and
   $w_1$ in the
flat CPL model from several popular SN Ia data sets which are
compiled with different light-curve fitters. ``Const" represents for
the Constitution SN Ia sample.}
\end{figure}

\begin{figure}[h!]
   \centering
       \includegraphics[width=0.45\linewidth]{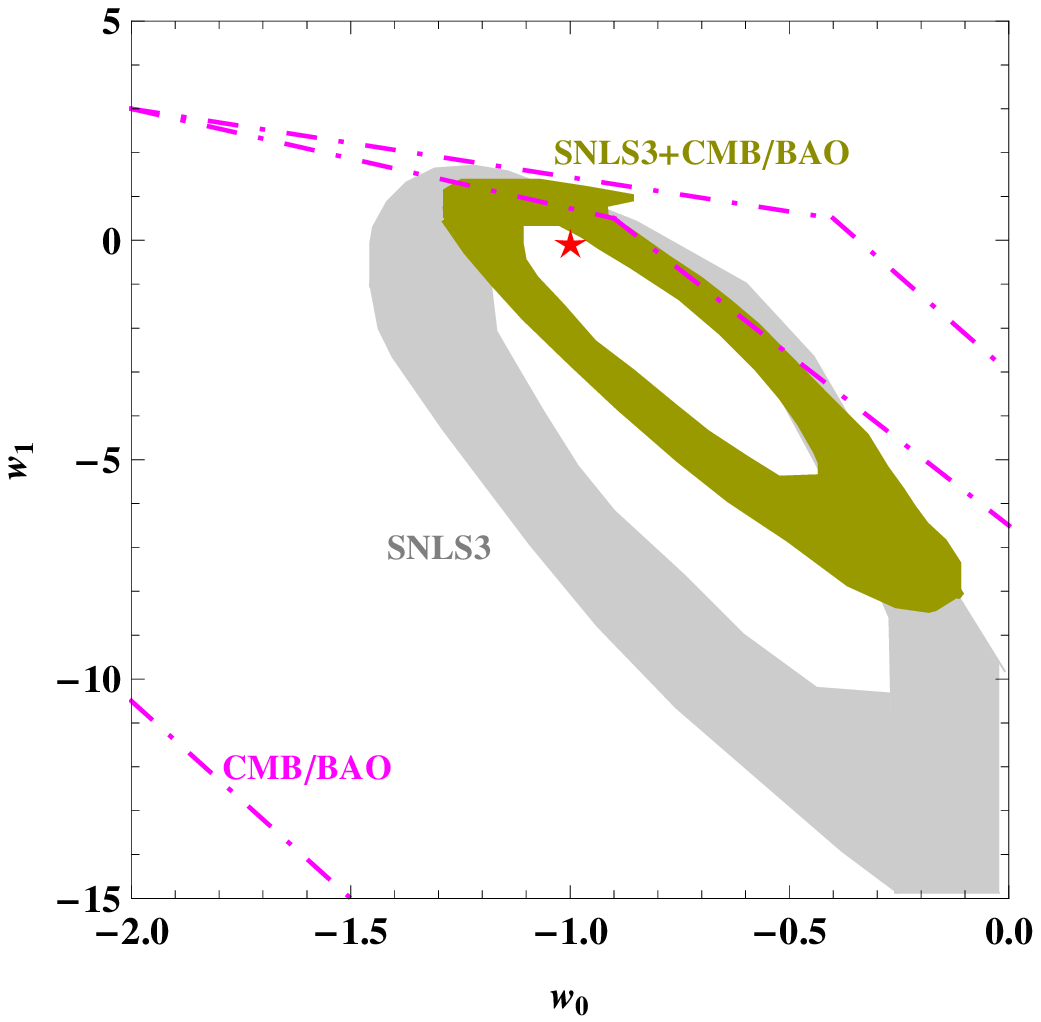}
       \includegraphics[width=0.45\linewidth]{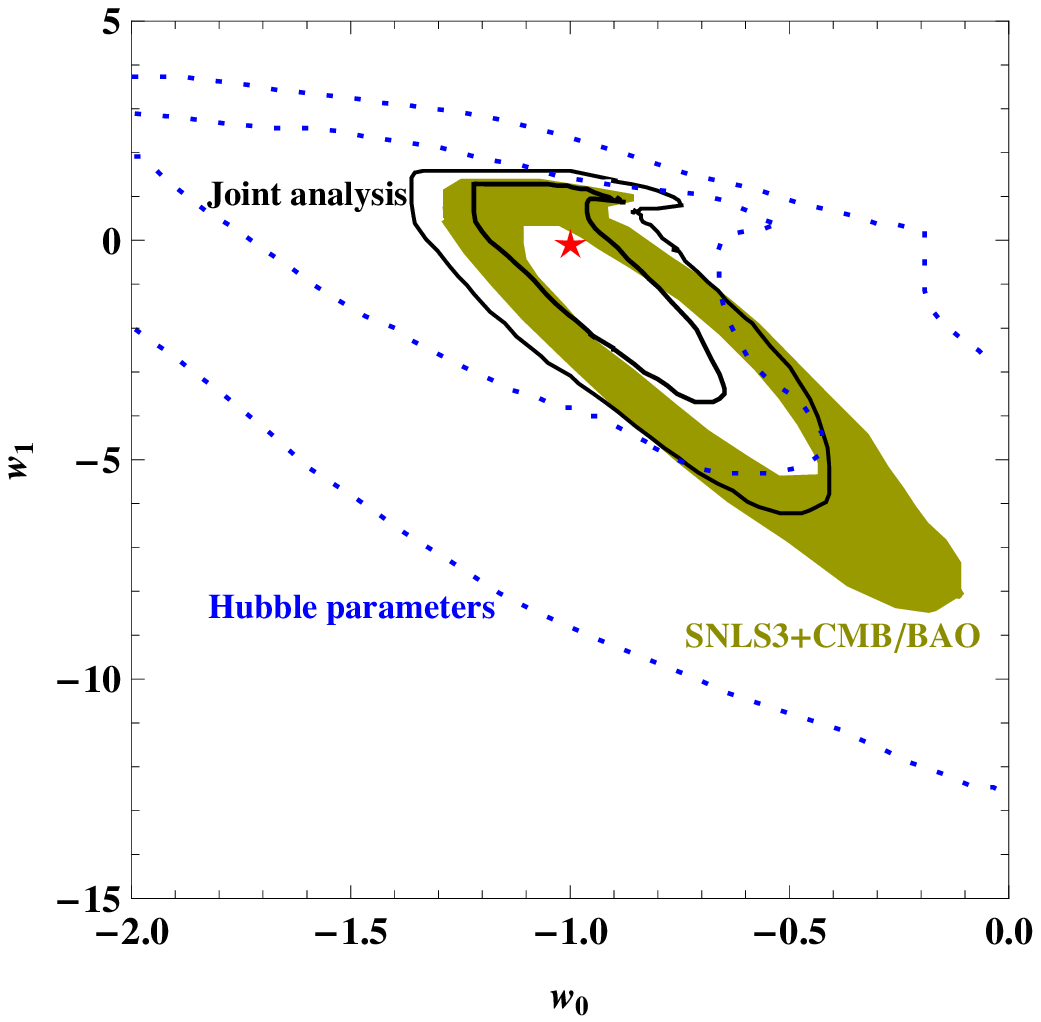}
   \caption{\label{FigC2} The marginalized $68.3\%$ and $95.4\%$ confidence level contours on $w_0-w_1$
   for the varying $w$ model with
the CPL parametrization. Left panel shows the constraints from
SNLS3, CMB/BAO and their combination,  and right panel  is the case
with the H(z) data included.
   The red star ($w_0=-1.0,~w_1=0.0$)
represents the $\Lambda$CDM.  }
\end{figure}

The degrees of freedom of the cosmic model increase when we allow
the dark energy equation of state to vary with the cosmic time. In
this case, the Friedmann equation for a  varying $w$ dark energy
model is given by
\begin{equation}
H^2=H_0^2\bigg\{\Omega_m(1+z)^3+(1-\Omega_m)\exp\bigg[3\int_0^z\frac{1+w(z')}{1+z'}dz'\bigg]\bigg\}.
\end{equation}
In this paper, we consider the popular CPL~\citep{CPL1, CPL2}
parametrization $w(z)=w_0+w_1z/(1+z)$. The above expression can then
be simplified to
\begin{equation}
H^2=H_0^2\bigg\{\Omega_{m}(1+z)^3+(1-\Omega_{m})(1+z)^{3(1+w_0+w_1)}\exp\bigg[-\frac{3w_1z}{1+z}\bigg]\bigg\}.
\end{equation}
This model has been investigated by using SNLS3 in S11 and
\citet{Miao}. In Fig.~\ref{FigC1}, we show the marginalized $68.3\%$
contours of $w_0$ and $w_1$ from different SN Ia data with different
light-curve fitters. Consistent results are obtained and the
tensions between different SN Ia sets and different light-curve
fitters occurring in the $\Lambda$CDM and the constant $w$ model
disappear when the CPL is considered. However,  this tension
improvements are probably simply due to the fact that this model is
not well constrained by current data sets, especially the $w_1$
constraints are so bad. A combination of SNLS3+BAO/CMB+H(z) gives
$\Omega_m=0.253^{+0.023}_{-0.020}$, $w_0=-1.007^{+0.187}_{-0.253}$,
and $w_1=-0.344^{+1.144}_{-2.656}$ at the $68.3\%$  confidence
level. The marginalized contours for $w_1$ and $w_0$ are plotted in
Fig.~\ref{FigC2}. One can see that the $\Lambda$CDM
($w_0=-1.0,~w_1=0.0$) is included at the $68.3\%$ confidence level
for all observational data sets, which are consistent with what
obtained in S11 from SNLS3+SDSS DR7 LRGs+WMAP7+$H_0$.

\subsection{Flat DGP model}
\begin{figure}[h!]
   \centering
       \includegraphics[width=0.8\linewidth]{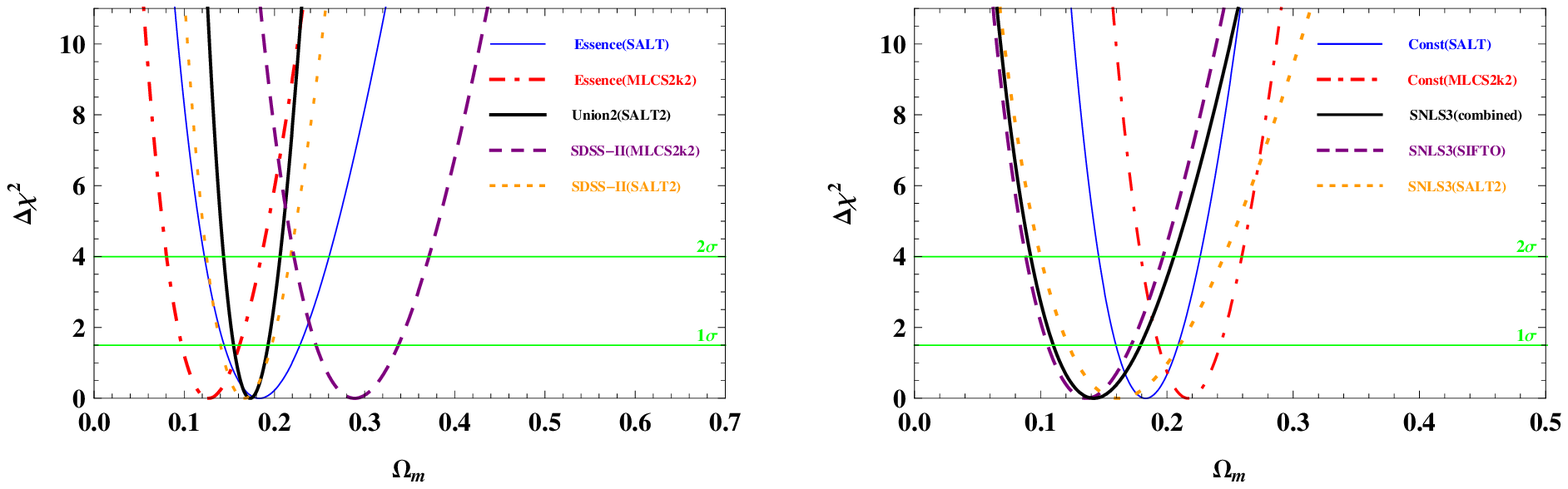}
   \caption{\label{FigD1}  The constraints on the
flat DGP model from several popular SN Ia data sets compiled with
different light-curve fitters. ``Const" represents for the
Constitution SN Ia sample.}
\end{figure}

\begin{figure}[h!]
   \centering
       \includegraphics[width=0.45\linewidth]{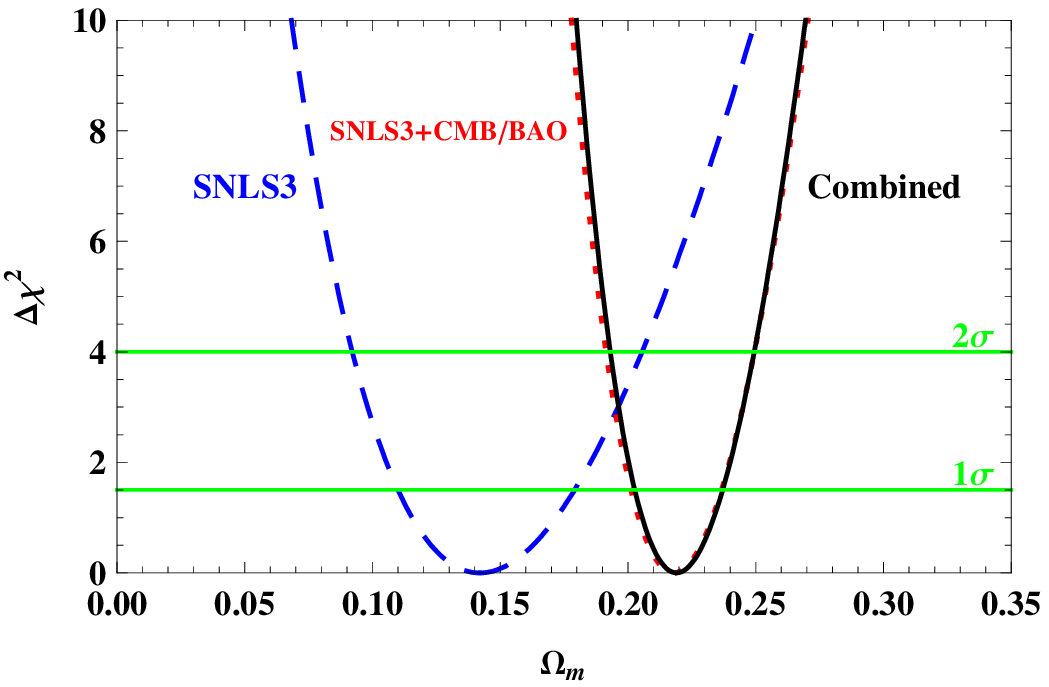}
   \caption{\label{FigD2} The constraints on the
   flat DGP model. The blue dashed, red dotted and solid lines
represent the results from SNLS3, SNLS3+CMB/BAO and
SNL3+CMB/BAO+H(z), respectively. }
\end{figure}

The DGP model~\citep{Dvali}, which accounts for the cosmic
acceleration without dark energy, arises from a class of brane world
theories in which gravity leaks out into the bulk at large
distances. For a spatially flat case, the Friedmann equation in the
DGP model can be expressed as
\begin{equation}
H^2=H_0^2[\sqrt{\Omega_m(1+z)^3+\Omega_{r_c}}+\sqrt{\Omega_{r_c}}]^2,
\end{equation}
where $\Omega_m=1-2\sqrt{\Omega_{r_c}}$. The parameter $r_c$
represents the critical length scale beyond which gravity leaks out
into the bulk and $\Omega_{r_c}$ is related to this critical length
by $\Omega_{r_c}=1/(4r_ch_0^2)$. The flat DGP model has the same
number of free parameter as the $\Lambda$CDM. The constraints from
different SN Ia samples are shown in Fig.~\ref{FigD1}. We obtain
similar results to the $\Lambda$CDM for the DGP model, except that
the latter favors a smaller $\Omega_m$. Fig.~\ref{FigD2} shows the
constraints on $\Omega_m$ from SNLS3, CMB/BAO and H(z) data, from
which we find that the SNLS3 gives a very small value for $\Omega_m$
and this value becomes large when the CMB/BAO data are added. The
combined SNLS3+CMB/BAO+H(z) gives
$\Omega_m=0.219^{+0.022}_{-0.020}$, which  is smaller than that
obtained in \citet{Xu, Liang} from other observations, where
$\Omega_m=0.297^{+0.037}_{-0.039}$ and $0.285^{+0.252}_{-0.066}$,
respectively.

It has been argued that the BAO data cannot be used to constrain the
DGP model, since the details of structure formation are unclear in
this model \citep{Rydbeck}. Moreover, it has been found that a
tension between distance measures and horizon scale growth in the
DGP exists and there is no way to alleviate it~\citep{Seahra, Song,
Fang}. With these caveats in mind,  we still present the CMB/BAO
constraints for the sake of completeness.

\subsection{Flat Modified Polytropic Cardassian}
\begin{figure}[h!]
   \centering
       \includegraphics[width=0.8\linewidth]{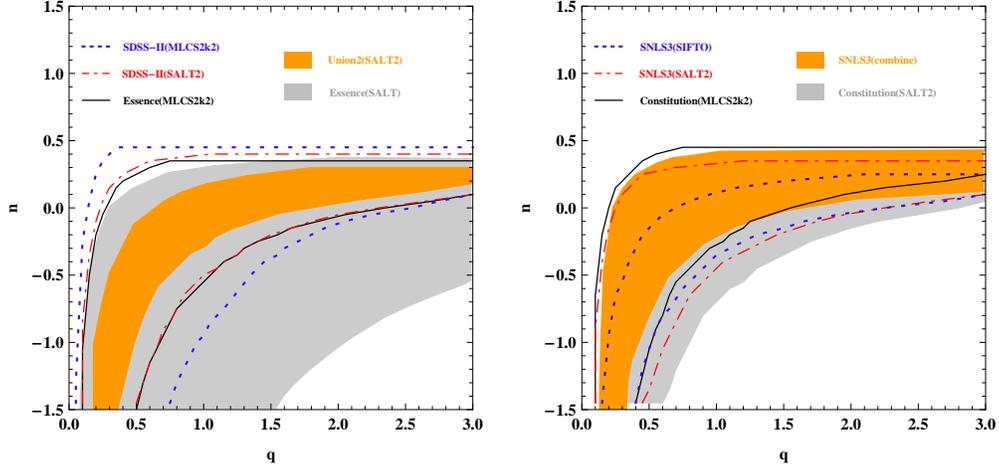}
   \caption{\label{FigCA1}  The marginalized $68.3\%$ contours for the
flat modified polytropic Cardassian model from several popular SN Ia
data sets which are compiled with different light-curve fitters. }
\end{figure}

\begin{figure}[h!]
   \centering
       \includegraphics[width=0.45\linewidth]{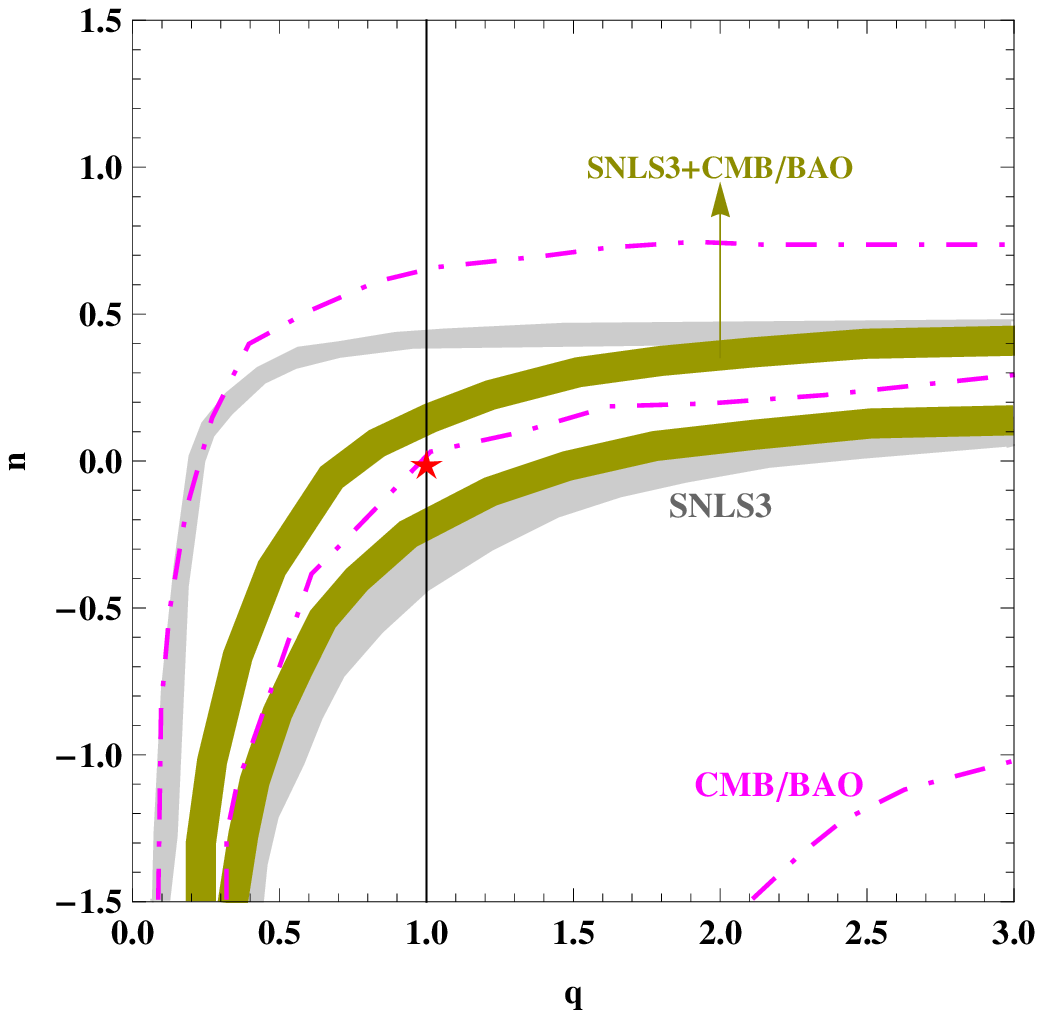}
       \includegraphics[width=0.45\linewidth]{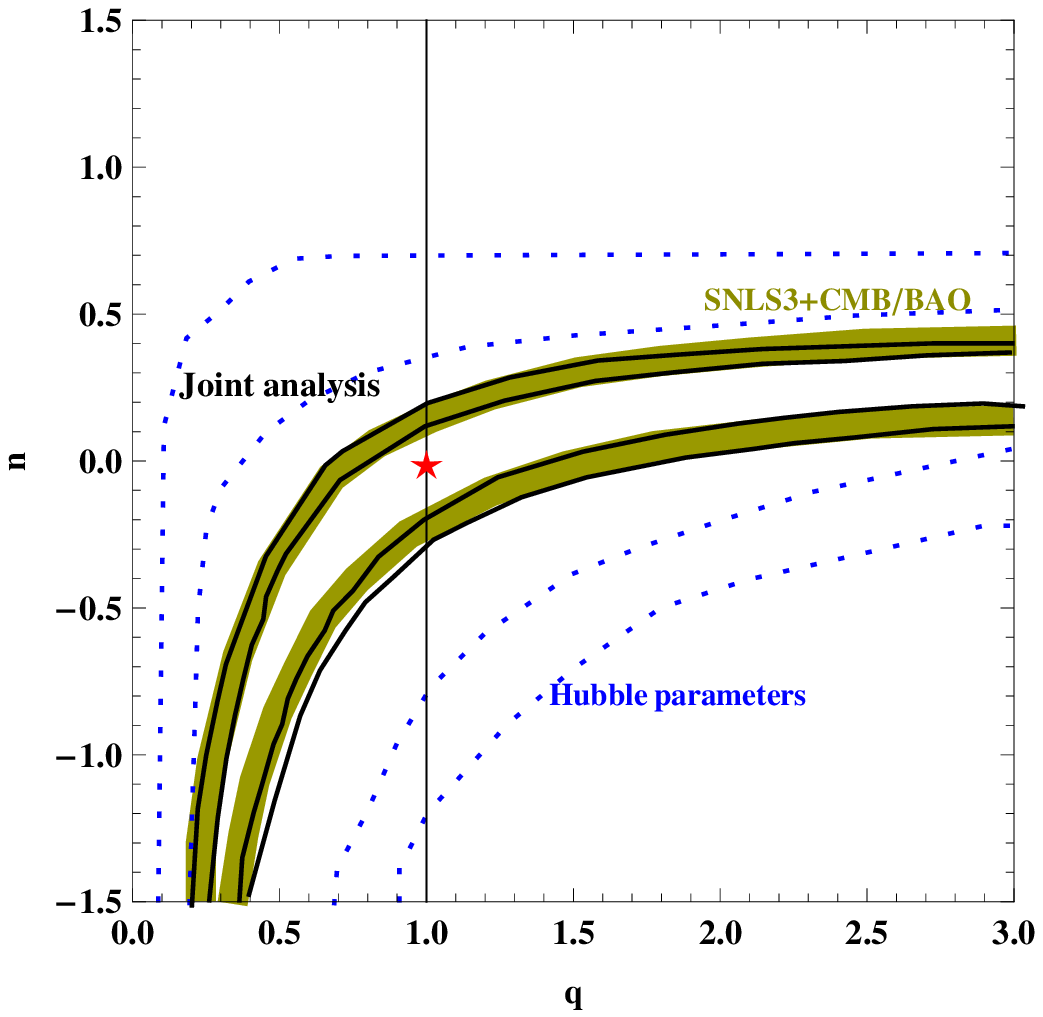}
   \caption{\label{FigCA2} The marginalized $68.3\%$ and $95.4\%$ confidence level contours  for
the flat modified polytropic Cardassian model. Left panel shows the
constraints from SNLS3, CMB/BAO and their combination,  and right
panel  is the case of  H(z) data included. The vertical line
($q=1.0$) represents the  constant $w(=n-1)$ model, and the red
   star ($q=1.0,~n=0.0$) corresponds to  the  $\Lambda$CDM one. }
\end{figure}

The Cardassian model~\citep{Freese} is based on a modified Friedmann
equation in which an additional term $B\rho_m^n$ is added to the
right hand side
\begin{equation}
H^2=\frac{8\pi G}{3}\rho_m+B\rho_m^n,
\end{equation}
where $\rho_m$ is the density of matter and $n$ is a dimensionless
parameter. This equation can be rewritten as
\begin{equation}
H^2=\frac{8\pi
G}{3}\rho_m\bigg[1+\bigg(\frac{\rho_{Card}}{\rho_m}\bigg)^{1-n}\bigg].
\end{equation}
This model reduces to the flat $\Lambda$CDM when $n=0$, and is
related to the constant $w$ model by $n=1+w$. Therefore, the results
for the original Cardassian model can be directly transcribed from
those of section~\ref{section 3.2} and there is no need of
additional fit for that model. Here, we consider a modified
Cardassian model, the modified polytropic Cardassian model, in which
the cosmic evolution is governed by~\citep{yunw}:
\begin{equation}
H^2=\frac{8\pi
G}{3}\rho_m\bigg[1+\bigg(\frac{\rho_{Card}}{\rho_m}\bigg)^{q(1-n)}\bigg]^{1/q}.
\end{equation}
The above expression can be reexpressed as
\begin{equation}
H^2=H_0^2\bigg\{\Omega_m(1+z)^3\big[1+(\Omega_m^{-q}-1)(1+z)^{3q(n-1)}\big]^{1/q}\bigg\}.
\end{equation}
$\Lambda$CDM is recovered  for $q=1$ and $n=0$. The marginalized
$68.3\%$ contours in the $q-n$ plane for different SN samples are
shown in Fig.~\ref{FigCA1}. As was the case for the varying $w$ CPL
model, here we find that the differences between the constraints
arising from different data  and different light-curve fitters are
negligible, which may also be due to the weak constraints on the
model parameters from these observations. The constraint results
from SNLS3 along with other data are shown in Fig.~\ref{FigCA2}. The
$\Lambda$CDM is well consistent with all observations  at the
$68.3\%$ confidence level. The combined analysis gives
$\Omega_m=0.248^{+0.047}_{-0.013}$, $q=1.098^{+1.015}_{-0.465}$ and
$n=0.014^{+0.364}_{-0.946}$, which is consistent with what given
in~\citet{Davis} (see Fig. 4) and  in~\citet{Wang}
($\Omega_m=0.271^{+0.014}_{-0.015}$, $q=0.824^{+0.750}_{-0.622}$,
$n=-0.091^{+0.331}_{-1.908}$).

\subsection{Flat Generalized Chaplygin Gas}
\begin{figure}[h!]
   \centering
       \includegraphics[width=0.8\linewidth]{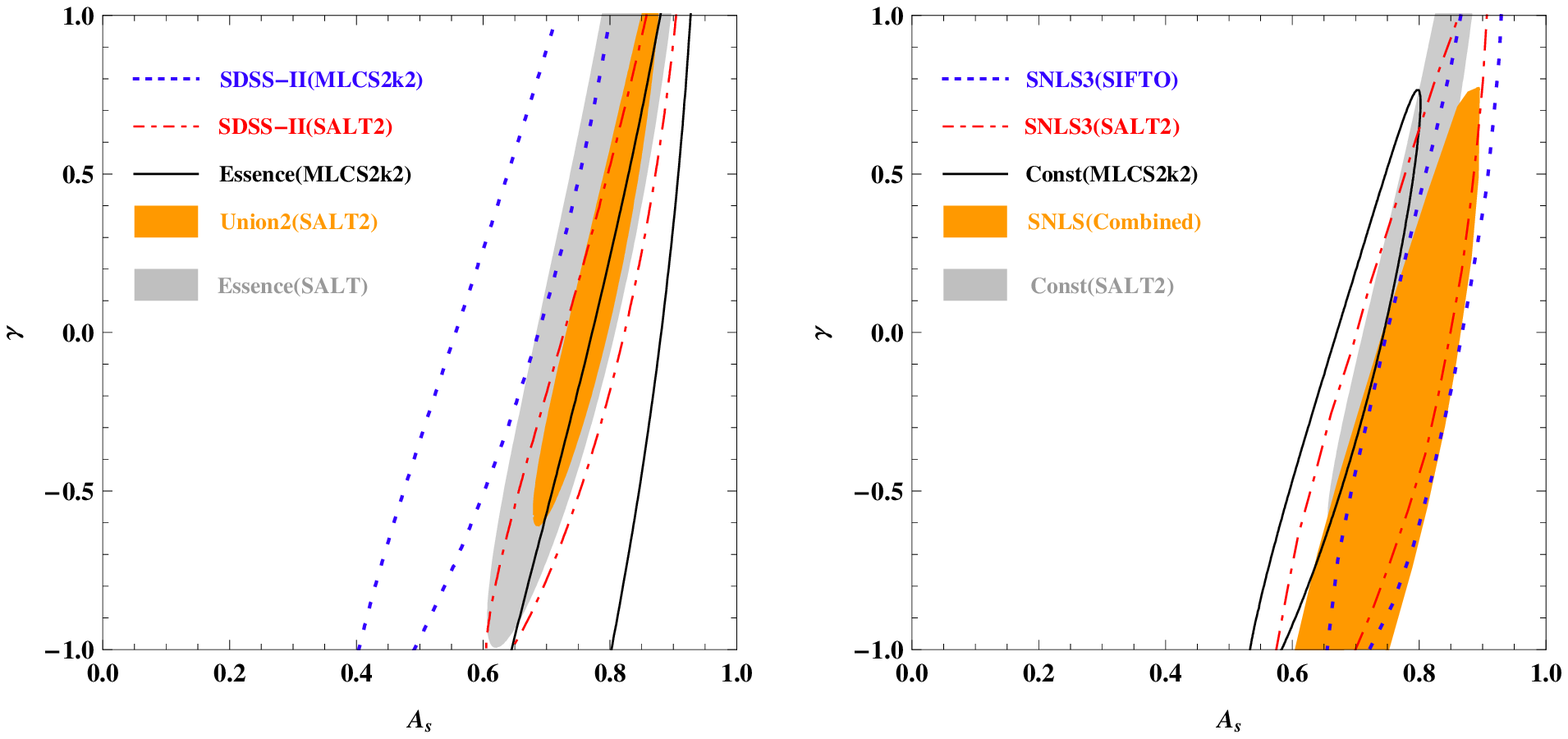}
   \caption{\label{FigG1}  The constraints on the
flat GCG model from several popular SN Ia data sets which are
compiled with different light-curve fitters.  ``Const" represents
for the Constitution SN Ia sample.}
\end{figure}

\begin{figure}[h!]
   \centering
       \includegraphics[width=0.45\linewidth]{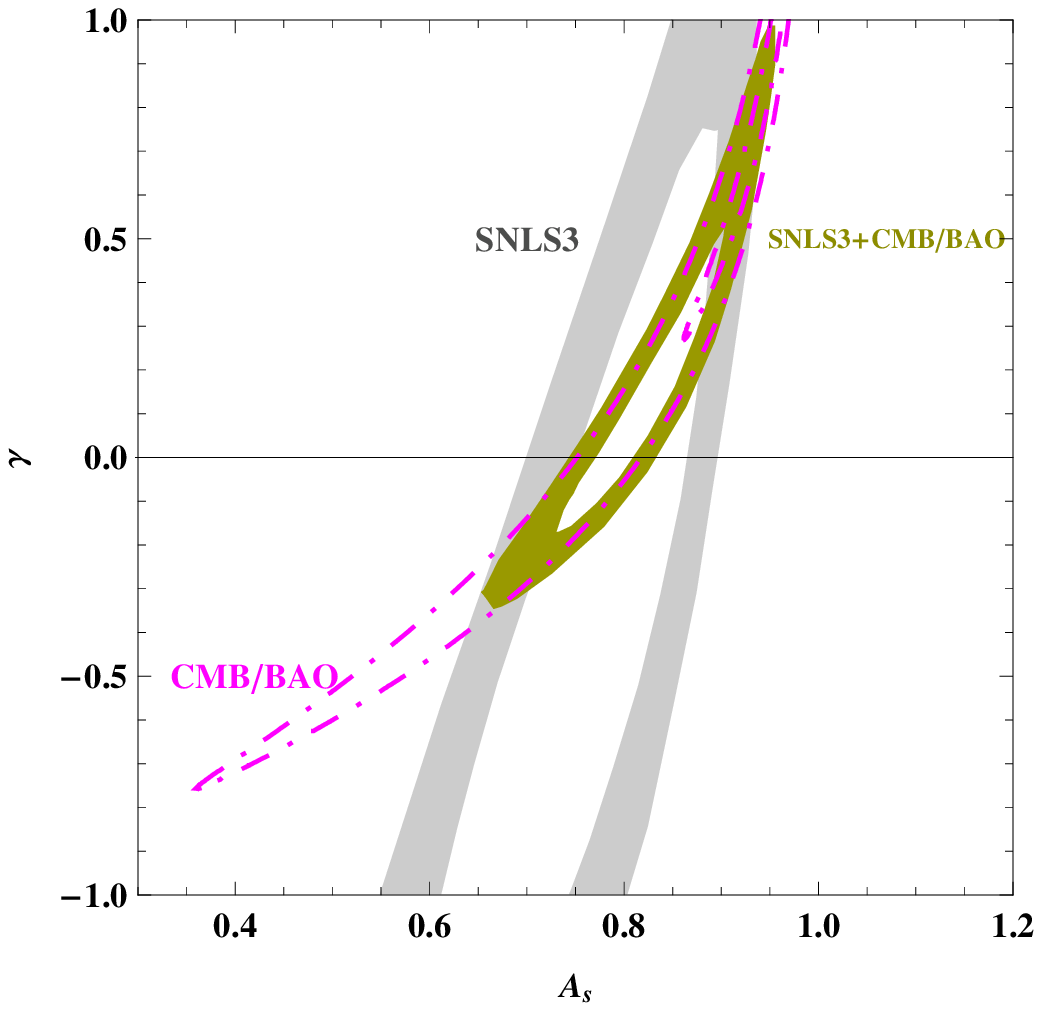}
       \includegraphics[width=0.45\linewidth]{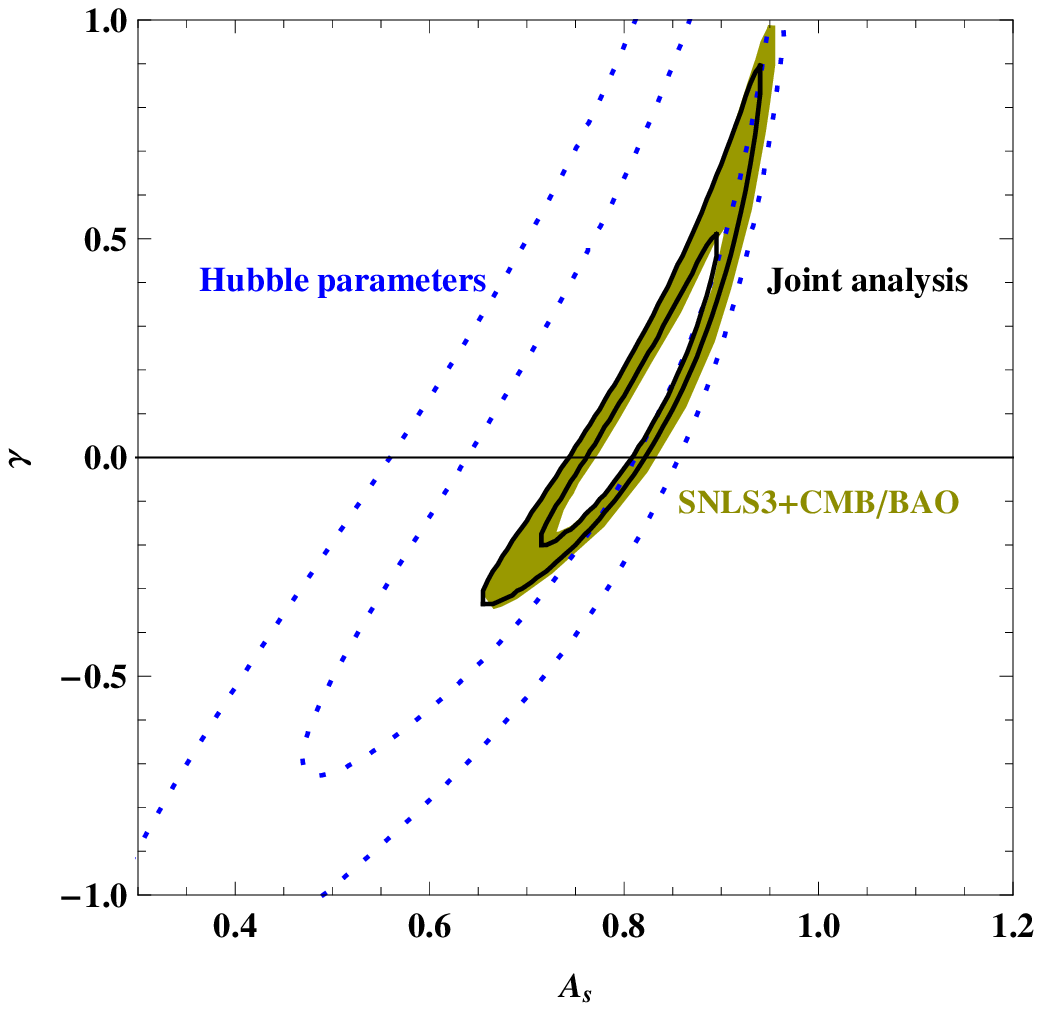}
   \caption{\label{FigG2} The $68.3\%$ and $95.4\%$ confidence level contours for
the flat GCG model. Left panel shows the constraints from SNLS3,
CMB/BAO and their combination, and right panel is the case with the
H(z) data included. The horizontal line ($\gamma=0.0$) represents
the standard cosmological constant model ($\Lambda$CDM) and
$\gamma=1.0$ corresponds to the original Chaplygin gas one. The
original Chaplygin gas model is ruled out by SNLS3+CMB/BAO and
SNLS3+CMB/BAO+Hubble at the $95.4\%$ confidence level, while, except
for the CMB/BAO data, the LCDM is allowed by observations at the
$68.3\%$ confidence level. }
\end{figure}

The Chaplygin gas model~\citep{Kamenshchik} unifies dark matter and
dark energy by invoking a background fluid with an equation of state
$p\propto\rho^{-\gamma}$, with $\gamma=1$ in the basic model. Here,
we consider a generalization of this model (the Generalized
Chaplygin Gas, or GCG model) by allowing $\gamma$ to take arbitrary,
but constant, values~\citep{Bento}. For the GCG, the Friedmann
equation is
\begin{equation}
H^2=H_0^2\big[\Omega_b(1+z)^3+(1-\Omega_b)\times\big(A_s+(1-A_s)(1+z)^{3(1+\gamma)})^{\frac{1}{1+\gamma}}\big],
\end{equation}
where $\Omega_b$ is the present dimensionless density parameter of
baryonic matter which is related to the effective matter density
parameter by $\Omega_m=\Omega_b+(1-\Omega_b)(1-A_s)^{1/(1+\gamma)}$
and the WMAP7 observation gives $\Omega_bh^2=0.02246$
~\citep{Komatsub}. Here $h$ is the Hubble constant in unit of
$100~\mathrm{Km}\cdot \mathrm{s}^{-1}\cdot \mathrm{Mpc}^{-1}$ and we
marginalize over it  by using the~\citet{Hubble2} $H_0$ prior. $A_s$
is a model parameter which relates the pressure $p$ and energy
density $\rho$ of the background fluid: $A_s=p/\rho^\gamma$.
$\gamma=0$ corresponds to the case of the cold dark matter plus a
cosmological constant. The $68.3\%$ contours  on $A_s$ and $\gamma$
from different SN data sets are shown in Fig.~\ref{FigG1}. The
results are similar to the constant $w$ model. Inconsistency appears
between SDSS-II (SALT2) and SDSS-II (MLCS2k2) and there is no
overlap between them at the $68.3\%$ confidence level. The
constraints from SNLS3 along with other cosmological probes are
shown in Fig.~\ref{FigG2}. By fitting this model to the combined
SNLS3 SN Ia, BAO/CMB and H(z) data, the constraints are
$A_s=0.810^{+0.085}_{-0.095}$ and $\gamma=0.086^{+0.434}_{-0.286}$.
We find that the standard Chaplygin gas model ($\gamma=1$) is ruled
out by  SNLS3+CMB/BAO and SNLS3+CMB/BAO+H(z) data at the $95.4\%$
confidence level, while the $\Lambda$CDM is consistent with them at
the $68.3\%$ confidence level. This is consistent with the results
of~\citet{Davis, Liang2, Lu, puxun, Wu1}.

\section{MODEL TESTING USING MODEL SELECTION STATISTICS}

In this section, we discuss the worth of models by applying model
comparison statistics such as $\chi^2/dof$ ($dof$: degrees of
freedom), goodness of fit (GoF), Akaike information
criterion~\citep{Akaike} and Bayesian information criterion
~\citep{Schwarz}. The $\chi^2/dof$   describes how well the model
fits a set of observations. The GoF simply gives the probability of
obtaining, by chance, a data set that is a worse fit to the model
than the actual data, assuming the model is correct. It
 is defined as
\begin{equation}
\mathrm{GoF}=\Gamma(\nu/2,~\chi^2/2)/\Gamma(\nu/2),
\end{equation}
where $\Gamma$ is the incomplete gamma function and $\nu$ is the
number of degrees of freedom. For a family of models, the best fit
one has a  minimum $\chi^2/dof$ value, while it has a maximum value
of  GoF.

For a given data set, the candidate models may be ranked according
to their AIC values, which can be calculated by
\begin{equation}\label{AIC}
\mathrm{AIC}=-2\ln \mathcal{L}+2k
\end{equation}
where $\mathcal{L}$ is the maximum likelihood, $k$ is the number of
model parameters. Given a set of candidate models for the data, the
one which has the minimum AIC value can be considered the best. Then
the relative strength of evidence for each model can be judged by
using the differences ($\Delta$AIC) between the AIC quantities of
the rest of models and that of the best one.  The models with
$0\leq\Delta\mathrm{AIC}\leq2$ are considered substantially
supported, those where $4\leq\Delta\mathrm{AIC}\leq7$ have less
support, while models with $\Delta\mathrm{AIC}>10$ are essentially
unsupported with respect to the best model~\citep{Szydlowski}.

The BIC, very similar to the AIC, is also a criterion for model
selection among a finite set of models. It is defined as
\begin{equation}
\mathrm{BIC}=-2\ln \mathcal{L}+k\ln N
\end{equation}
where $N$ is the number of data points used in the fit. The model
which minimizes the  BIC is the best fit one. As the AIC,    the
differences between the BIC of the rest of models and that of the
best one ($\Delta$BIC) is used for  the judgement of the model, that
is, $0\leq\Delta\mathrm{BIC}\leq2$ is considered  as a weak,
$2\leq\Delta\mathrm{BIC}\leq6$ as a positive,
$6\leq\Delta\mathrm{BIC}\leq10$ as a strong and
$\Delta\mathrm{BIC}>10$ as a very strong evidence favoring a better
model~\citep{Liddle, Szydlowski}. Both the AIC and BIC penalize the
case of adding the model parameter to increase the likelihood
through the introduction of  a penalty term, which depends on the
number of parameters in the model. But the coefficients in this
penalty term are different for the AIC and BIC. It should be noticed
that  in the limit of large data (large N) the AIC tends to favor
models with more parameters while the BIC tends to penalize
them~\citep{Parkinson}.

The results for these model tests are shown in
Table~\ref{Tab2},~\ref{Tab3},~\ref{Tab4}. From these tables, we find
that, the DGP model is favored by the SNLS3 data alone. However, if
additional constraints such as CMB/BAO and H(z) are included, this
is no longer true, and the DGP model is disfavored by most criteria.
Instead, the flat $\Lambda$CDM becomes preferred. The GCG and F$w$
are hardly distinguishable by most selection statistics and they
become the favored  non-standard models (excluding $\Lambda$CDM)
when the CMB/BAO and H(z) are added. The variable $w$ (CPL) and
Cardassian (MPC) models are penalized the most because of their
large number of cosmological parameters.

\section{Conclusion}
In this paper, we compare the constraints from the enlarged first
three years data of the Supernova Legacy Survey (SNLS3) with those
from other SN Ia samples. Several popular SN Ia samples, such as
ESSENCE, Union2, Constitution and SDSS-II are used and the effects
of different light-curve fitters on results are considered.  We also
discuss the observational constraints from the SNLS3, together with
other two cosmological probes, CMB/BAO and Hubble parameter versus
redshift. The $\Lambda$CDM and five nonstandard cosmological models
are considered.  We find that, for models with fewer  free
parameters (F$\Lambda$, FDGP, F$w$ and FGCG; see Table~\ref{Tab1}),
different SN Ia samples give fairly consistent results when analyzed
with the SALT2, SALT, or SIFTO light-curve fitters, but this is not
true when the MLCS2k2 light-curve fitter is used. Moreover, we find
significant tension between SDSS-II sample results when analyzed
with SALT2 or MLCS2k2, as noted in~\citet{Sollerman}. The
inconsistencies between different samples and different light-curve
fitters seem to be reduced for models with more free parameters
(FCPL and FMPC). This improvement may be ascribed to the fact that
current observational data has a weak constraint on the model
parameters.

By combining the SNLS3 with CMB/BAO and Hubble parameter versus
redshift data, we find that, except for the flat varying $w$ model,
the main constraints on model parameters come from the SNLS3+CMB/BAO
and the H(z) data has little effect. Moreover, we study the worth of
models by applying model comparison statistics such as $\chi^2/dof$,
goodness of fit (GoF), Akaike information criterion and Bayesian
information criterion. We find that the DGP model is the best one
for SNLS3 alone while the flat $\Lambda$CDM is preferred when the
CMB/BAO or H(z) data is included. When only the nonstandard models
are considered ($\Lambda$CDM is excluded),  the GCG and F$w$ are
preferred by both SNLS3+CMB/BAO and   SNLS3+CMB/BAO+H(z) data. These
results are independent of the test methods, that is, different test
methods give consistent conclusions. Our results are slightly
different from that of the analysis of the SDSS-II~\citep{Sollerman}
and ESSENCE SN Ia data~\citep{Davis}. In \citet{Sollerman}, both
SDSS-II (MLCS2k2) and SDSS-II (MLCS2k2) plus CMB/BAO prefer the DGP
model, while, once the SDSS-II (SALT2) is used, the $\Lambda$CDM is
always the best fit one.  In \citet{Davis}, the combination of
ESSENCE (MLCS2k2), BAO, and CMB  favors the $\Lambda$CDM. Finally,
the FCPL and FMPC suffer in BIC test due to the extra model
parameters, which agrees with the conclusions
in~\citep{Davis,Sollerman}.

\acknowledgments We would like to thank A. Conley for the insightful
comments and very helpful suggestions. This work was supported by
Hunan Provincial Innovation Foundation For Postgraduate, the
National Natural Science Foundation of China under Grants Nos.
10935013, 11175093 and 11075083, Zhejiang Provincial Natural Science
Foundation of China under Grants Nos. Z6100077 and R6110518, the
FANEDD under Grant No. 200922, the National Basic Research Program
of China under Grant No. 2010CB832803, the NCET under Grant No.
09-0144,  the PCSIRT under Grant No. IRT0964, the Hunan Provincial
Natural Science Foundation of China under Grant No. 11JJ7001, and
the Program for the Key Discipline in Hunan Province.

\begin{table}[!h]
\begin{center}
\begin{tabular}{l}
\toprule[1.5pt]
Model~~~~~~~~~~~~~~~~~~~~~~$\chi^2/dof$~~~~~~~~~~~~~~~~~~GoF(\%)~~~~~~~~~~~$\Delta$AIC~~~~~~~~~~~~$\Delta$BIC\\
\midrule[0.5pt]
FDGP.....~~~~~~~~~~419.464/472(0.8887)~~~~~~~~~~~96.04~~~~~~~~~~~~~~0.00~~~~~~~~~~~~~~0.00\\
F$\Lambda$.........~~~~~~~~~~~419.691/472(0.8892)~~~~~~~~~~~95.98~~~~~~~~~~~~~~0.23~~~~~~~~~~~~~~0.23\\
F$w$.........~~~~~~~~~~~419.282/471(0.8902)~~~~~~~~~~~95.81~~~~~~~~~~~~~~1.82~~~~~~~~~~~~~~5.98\\
FGCG....~~~~~~~~~~~419.342/471(0.8903)~~~~~~~~~~~95.79~~~~~~~~~~~~~~1.88~~~~~~~~~~~~~~6.04\\
FMPC....~~~~~~~~~~~418.472/470(0.8904)~~~~~~~~~~~95.77~~~~~~~~~~~~~~3.01~~~~~~~~~~~~~~11.32\\
FCPL.....~~~~~~~~~~~418.664/470(0.8908)~~~~~~~~~~~95.71~~~~~~~~~~~~~~3.20~~~~~~~~~~~~~~11.51\\
\bottomrule[1.5pt]
\end{tabular}
\end{center}
\caption{\label{Tab2} Summary of the model test results for the
SNLS3. The models are listed in the increasing order of the
$\Delta$AIC values (please refer to Table~\ref{Tab1}  for the
definitions of the model acronyms). When only the SNLS3 is
considered, the flat DGP is preferred. }
\end{table}

\begin{table}[!h]
\begin{center}
\begin{tabular}{l}
\toprule[1.5pt]
Model~~~~~~~~~~~~~~~~~~~~~~$\chi^2/dof$~~~~~~~~~~~~~~~~~~GoF(\%)~~~~~~~~~~~$\Delta$AIC~~~~~~~~~~~~$\Delta$BIC\\
\midrule[0.5pt]
F$\Lambda$.........~~~~~~~~~~~422.741/474(0.8919)~~~~~~~~~~~95.61~~~~~~~~~~~~~~0.00~~~~~~~~~~~~~~0.00\\
FGCG....~~~~~~~~~~~422.487/473(0.8932)~~~~~~~~~~~95.38~~~~~~~~~~~~~~1.75~~~~~~~~~~~~~~5.91\\
F$w$.........~~~~~~~~~~~422.603/473(0.8935)~~~~~~~~~~~95.34~~~~~~~~~~~~~~1.86~~~~~~~~~~~~~~6.02\\
FCPL.....~~~~~~~~~~~421.124/472(0.8922)~~~~~~~~~~~95.52~~~~~~~~~~~~~~2.38~~~~~~~~~~~~~~10.71\\
FMPC....~~~~~~~~~~~421.213/472(0.8924)~~~~~~~~~~~95.49~~~~~~~~~~~~~~2.47~~~~~~~~~~~~~~10.79\\
FDGP.....~~~~~~~~~~~429.995/474(0.9072)~~~~~~~~~~~92.71~~~~~~~~~~~~~7.25~~~~~~~~~~~~~~~7.25\\
\bottomrule[1.5pt]
\end{tabular}
\end{center}
\caption{\label{Tab3} Summary of the model test results for
SNLS3+CMB/BAO. The models are listed in the increasing order of the
$\Delta$AIC values (please refer to Table~\ref{Tab1}  for the
definitions of the model acronyms). When the additional CMB/BAO is
included, the flat DGP is disfavored and the flat $\Lambda$CDM
becomes preferred. }
\end{table}

\begin{table}[!h]
\begin{center}
\begin{tabular}{l}
\toprule[1.5pt]
Model~~~~~~~~~~~~~~~~~~~~~~$\chi^2/dof$~~~~~~~~~~~~~~~~~~GoF(\%)~~~~~~~~~~~$\Delta$AIC~~~~~~~~~~~~$\Delta$BIC\\
\midrule[0.5pt]
F$\Lambda$.........~~~~~~~~~~~436.106/488(0.8937)~~~~~~~~~~~95.57~~~~~~~~~~~~~~0.00~~~~~~~~~~~~~~0.00\\
FGCG....~~~~~~~~~~~435.931/487(0.8951)~~~~~~~~~~~95.31~~~~~~~~~~~~~~1.83~~~~~~~~~~~~~~6.02\\
F$w$.........~~~~~~~~~~~435.938/487(0.8951)~~~~~~~~~~~95.31~~~~~~~~~~~~~~1.83~~~~~~~~~~~~~~6.02\\
FMPC....~~~~~~~~~~~435.656/486(0.8964)~~~~~~~~~~~95.08~~~~~~~~~~~~~~3.55~~~~~~~~~~~~~~11.93\\
FCPL.....~~~~~~~~~~~435.900/486(0.8969)~~~~~~~~~~~94.99~~~~~~~~~~~~~~3.79~~~~~~~~~~~~~~12.18\\
FDGP.....~~~~~~~~~~~443.030/488(0.9078)~~~~~~~~~~~92.85~~~~~~~~~~~~~~6.92~~~~~~~~~~~~~~~6.92\\
\bottomrule[1.5pt]
\end{tabular}
\end{center}
\caption{\label{Tab4} Summary of the model test results for
SNLS3+CMB/BAO+H(z). The models are listed in the increasing order of
the $\Delta$AIC values (please refer to Table~\ref{Tab1}  for the
definitions of the model acronyms). When the additional H(z) is
included, the flat DGP is still disfavored and the flat $\Lambda$CDM
remains preferred.}
\end{table}

\end{document}